\documentstyle[12pt,preprint,aps]{revtex}%
\begin{document}
\title
{Quasi-Homogeneous Dynamical Structure Factor for Atomic-Trap Bose Condensates} 

\author
{Eddy Timmermans and Paolo Tommasini}
\address{Institute for Theoretical Atomic and Molecular Physics}
\address{Harvard-Smithsonian Center for Astrophysics}
\address{Cambridge, MA 02138}

\date{\today}
\maketitle
\begin{abstract}

	The essence of the Thomas-Fermi model is the assumption that the local
behavior of a many-body system can be approximated by that of a homogeneous
system.  In this paper, we present the natural extension of the 
static Thomas-Fermi
treatment of dilute Bose condensates, by describing the dynamical behavior
of the condensate in the same quasi-homogeneous approximation.
In particular, we calculate the dynamical structure factor 
$S({\bf q},{\bf \omega})$ of a low-temperature condensate, 
confined in a harmonic oscillator
trap.  The result is a remarkably simple analytical function, 
which, with the proper interpretation, gives a powerful and insightful 
description
of the scattering properties of the BEC-system.
 
\end{abstract}


\narrowtext

\section{Introduction}
	The achievement \cite{Ket}--\cite{Hul} of the long-elusive 
\cite{trends} goal of atomic Bose-Einstein condensation has set the
stage for exciting applications, such as tests of fundamental mean-field
theories \cite{proc} and the atom laser \cite{zol}.
The engineering problems associated with most 
applications are generally
less severe if the condensate is highly populated and it might be
necessary to create `larger' condensates before a practical use of 
atomic-trap BEC
can be demonstrated.  Regarding this point, we note
that recent experiments (see for example Ref. \cite{Ketlat})
report a significant increase in number of condensed atoms. 
Thus, the `large condensate' or `Thomas-Fermi' limit,
aside from its convenient simplicity,
is also of great importance from a practical point of view.

	The essence of the Thomas-Fermi model
is the `quasi-homogeneous picture', which is
the assumption that the behavior of a condensate near a position ${\bf r}$,
can be approximated by that of a homogeneous BEC 
with chemical potential equal to the local effective chemical potential:
\begin{equation}
\mu ({\bf r}) = \mu_{T} - V({\bf r}),
\label{e:mueff}
\end{equation}
where $\mu_{T}$ is the chemical potential of the trapped condensate and
$V({\bf r})$ the trapping potential. 
Although many authors \cite{Leg}--\cite{hung} have explored the physics
and the limits of this description, most efforts have been limited to
the static description of a dilute BEC.
In fact, the few studies that have calculated dynamical properties
\cite{wu}--\cite{hung} starting from a 
Thomas-Fermi description, have gone beyond
the quasi-homogeneous picture in order 
to obtain a discrete spectrum.  This allowed the description of interesting
effects due to the finite size of the system, 
but caused the formalism to lose some
of the simplicity of the straightforward quasi-homogeneous description.  
Perhaps, the inability
of the quasi-homogeneous description to calculate such finite-size effects
is to be blamed for the lack in the literature of a simple 
quasi-homogeneous treatment of the dynamical BEC structure.

	In this paper, we present such treatment, and
investigate the limits and restrictions inherent to the
model. The simplicity of the results, combined
with a proper understanding of the limits, gives a powerful formalism 
that can be used to interpret experimental data, as well as to check
certain limits of more complicated computational schemes.  Furthermore,
the insight gained from studying the limits of validity, indicates
how experimental data taken from a finite-size system can be interpreted
in terms of the homogeneous system. 
 
	The quantity that we choose to calculate is the dynamical structure 
factor, which 
represents the information content of non-resonant scattering data
about the dynamical
many-body structure of the scattering system.
More relevant to the atomic-trap BEC-systems, resonant light scattering
gives a cross-section which in the off-resonant limit is also
proportional to the dynamical structure factor \cite{Jav1}, \cite{Jav2}, 
\cite{usl}.
More precisely, the single-scattering 
differential cross section $d^{2} \sigma /
d\Omega  d\omega$, where $d\Omega$ is an infinitesimal solid angle
and $\omega$ the energy transfer ($\hbar = 1$ in our units), is equal to
\begin{equation}
\frac{d^{2} \sigma}{d \Omega\; d\omega}
= |f({\bf q})|^{2}  S({\bf q},\omega)\; ,
\label{e:lvh}
\end{equation}
\noindent
where ${\bf q}$ is the momentum transfer, $S({\bf q},\omega)$ 
the dynamical structure factor of the many-body scattering system,
and $f({\bf q})$ the scattering
length that describes the scattering
of an incident particle by an individual target particle.
For non-resonant scattering, the scattering length can depend
explicitly on the momentum transfer, whereas for off-resonant light scattering
the scattering length is the large-detuning limit of the usual
single-atom resonant scattering length ($\sim \lambdabar \gamma /\Delta$, where
$\lambdabar$ is the inverse of the resonant wave number, $\gamma$ the width
of the excited resonant state and $\Delta$ the detuning of the incident
light).

	The structure factor
is the Fourier-transform of the density-density correlation function,
\begin{equation}
S({\bf q},\omega) = (2 \pi)^{-1} \int d^{3} x \; d^{3} x' dt'
\; \exp \left[-i {\bf q} \cdot \left( {\bf x}'-{\bf x} \right)
-\omega t' \right]
\; \langle \hat{\rho} ({\bf x}',t') \hat{\rho} ({\bf x},0) \rangle \; \; ,
\label{e:sf}
\end{equation}
where $\hat{\rho}$ represents the density operator and
$\langle \; \rangle$ denotes the thermally
averaged expectation value.
A substitution of the spatial integration variables in (\ref{e:sf})
to sum and difference variables, 
${\bf R} = \left[ {\bf x} + {\bf x}' \right] /2$
and ${\bf r} = {\bf x}'-{\bf x}$, gives
\begin{eqnarray}
&& \; \; \; \; \;
S({\bf q},\omega ) = \int d^{3} R  \; \sigma ({\bf R};{\bf q},\omega )
\nonumber \\
&& \rm{where} \; \;  
\sigma ({\bf R};{\bf q},\omega ) = (2 \pi )^{-1}
\int d^{3} r \; dt' \;  
\exp \left[ -i( {\bf q} \cdot {\bf r} - \omega t') \right]
\langle \hat{\rho} ({\bf R}+{\bf r}/2,t') \hat{\rho} ({\bf R}-{\bf r}/2,0)
\rangle
\label{e:sfd}
\end{eqnarray}
is a dynamical structure factor density. 
In the quasi-homogeneous approximation, we replace the correct structure
density by $\sigma^{({\rm hom})}_{\mu ({\bf R})} ({\bf q},\omega )$,
its value for a homogeneous system of 
chemical potential $\mu ({\bf R})$:
\begin{equation}
S({\bf q},\omega ) \approx S_{TF} ({\bf q},\omega) = 
\int d^{3} R \; \sigma^{({\rm hom})}_{\mu ({\bf R})} ({\bf q},\omega )\;
\; \; \; ,
\label{e:tfs}
\end{equation}
which leads to an analytical expression, as we show below. 

\section{Homogeneous BEC in the Bogoliubov approximation}
	In treating a homogeneous system, the natural choice for a 
single-particle basis is the set of plane wave states, labeled by their
wave vector/momentum ${\bf k}$.
Describing the interparticle interaction in the shape-independent
approximation by means of a scattering length a, the second-quantized
Hamiltonian operator reads
\begin{equation}
\hat{H} = \sum_{\bf k} (k^{2}/2m - \mu ) c^{\dagger}_{\bf k} c_{\bf k}
+ \frac{\lambda}{2V} \sum_{{\bf k},{\bf k}',{\bf q}}
c^{\dagger}_{{\bf k}+{\bf q}} c^{\dagger}_{{\bf k}'-{\bf q}} c_{{\bf k}'}
c_{\bf k}
\label{e:h}
\end{equation}
where we have included the chemical potential $\mu$, 
$\lambda = 4 \pi a/m$, V denotes the macroscopic volume of the 
system, and $c,c^{\dagger}$ represent the annihilation and creation operators.

	For the purpose of describing the static properties of a BEC 
of N atoms with a coherent condensate of an average of $N_{0}$ atoms 
in the ${\bf k} = 0$-mode, we can replace
$c_{{\bf k}=0}$ and $c^{\dagger}_{{\bf k}=0}$ by $\sqrt{N_{0}}$.
Provided the depletion is
low, $(N-N_{0})/N \ll 1$, 
we can neglect terms that contain less than two 
factors of $\sqrt{N_{0}}$. In this approximation, only valid for dilute
($na^{3} \ll 1$) and low temperature Bose condensates 
($k_{B} T < \lambda n_{0}$, where $n_{0}$ is the condensate density, 
$n_{0} = N_{0}/V$, T 
the temperature and $k_{B}$ the Boltzmann constant),
the hamiltonian is phonon-like:
\begin{eqnarray}
\hat{H} \approx V \; (-\mu n_{0} + \frac{\lambda}{2} n_{0}^{2} )
+ \sum_{{\bf k} \neq 0} (k^{2}/2m - \mu + n_{0} \lambda )
c^{\dagger}_{\bf k} c_{\bf k}  \; +
\nonumber\\
n_{0} \lambda \sum_{{\bf k} \neq 0} \; 
\frac{\left(c^{\dagger}_{\bf k} + c_{-{\bf k}}\right)}{\sqrt{2}} \; 
\frac{\left(c^{\dagger}_{-{\bf k}} + c_{\bf k}\right)}{\sqrt{2}}
\; \; \; \; \; . 
\label{e:phonh}
\end{eqnarray}
The free energy of the system, $F$, also contains the entropy, which we 
denote by S, and $F$ is equal to \cite{Paolo} 
$F = \langle \hat{H} \rangle - T S$.

	To lowest order, consistent with the assumption of 
low depletion, we can approximate
the expectation value of $\hat{H}$ by the first term of (\ref{e:phonh})
\begin{equation}
\langle \hat{H} \rangle \approx V \; (-\mu n_{0} + \frac{\lambda}{2}
n_{0}^{2} ) \; \; ,
\label{e:h1}
\end{equation}
and determine the chemical potential by minimizing $F$
with respect to $n_{0}$, giving 
\begin{equation}
\mu \approx n_{0} \lambda \; .
\label{e:mu}
\end{equation} 
To determine expectation values involving non-zero wavenumbers, we 
return to (\ref{e:phonh}) with (\ref{e:mu}). In doing so, we also find it 
useful to introduce Hermitian operators
\begin{eqnarray}
\Phi_{\bf k} = \frac{\left(c_{-{\bf k}} + c^{\dagger}_{\bf k}\right)}
{\sqrt{2}} \; \; ,
\nonumber \\
\Pi_{\bf k} = \frac{\left(c_{\bf k}-c^{\dagger}_{-{\bf k}}\right)}
{i \sqrt{2}} \; \; ,
\label{e:fluct}
\end{eqnarray}
which represent the fluctuations of the density and velocity field
\cite{LL}.  The canonical boson commutator relations such as
$\left[ c^{\dagger}_{\bf k},c_{{\bf k}'} \right] = \delta_{{\bf k},{\bf k}'}$,
are equivalent to the requirement that $\Phi_{\bf k}$ and $\Pi_{\bf k}$ are
canonically conjugate: $\left[ \Phi_{\bf k},\Pi_{{\bf k}'} \right]
= i \delta_{{\bf k},{\bf k}'}$.
With the fluctuation operators the Hamiltonian takes on the following form:
\begin{equation}
\hat{H} = \sum_{\bf k} (k^{2}/2m) \frac{1}{2} (\Phi_{\bf k} \Phi_{-{\bf k}}
+ \Pi_{-{\bf k}} \Pi_{\bf k}) + \mu \Phi_{\bf k} \Phi_{-{\bf k}}
\; \; \; \; . 
\label{e:hamfluc}
\end{equation}
In obtaining (\ref{e:hamfluc}), we have neglected a 
constant term, unimportant in describing the behavior 
of the system.

	At this point, we invoke the quasi-particle concept and introduce
quasi-particle fluctuation operators $\Phi^{'}_{\bf k}, \Pi^{'}_{\bf k}$,
defined as in (\ref{e:fluct}) with the particle creation and annihilation
operators replaced by quasi-particle creation and
annihilation operators.  We cast the Bogoliubov transformation between particle 
and quasi-particle operators in terms of the fluctuation
operators.  It can be seen that this transformation 
cannot mix $\Pi$ and $\Phi$ operators because
of time reversal symmetry.  Indeed, the density fluctuation expectation
value of a time-reversed state is equal to the expectation value
of the state, whereas the expectation value of the velocity-field fluctation 
changes sign.  The only transformation that preserves canonicity is then
a simple `scaling' transformation:
\begin{eqnarray}
&& \Phi_{\bf k} = \; \alpha_{\bf k} \; \; \Phi^{'}_{\bf k} \; \; ,
\nonumber \\
&& \Pi_{\bf k}= \; \alpha^{-1}_{\bf k} \; \Pi^{'}_{\bf k} \; \; ,
\label{e:scal}
\end{eqnarray}
which is the Bogoliubov transformation in fluctuation
operator notation.
We choose $\alpha$ real so that all $\Phi$ and $\Pi$-operators
are Hermitian.  In addition, the isotropical nature of the system suggests that
the scaling parameter $\alpha_{\bf k}$ 
should only depend on the magnitude of the momentum,
$\alpha_{\bf k} = \alpha_{k}$.  
The value of the scaling parameter is determined by
minimizing the free energy with respect to $\alpha_{k}$. In computing the
expectation value of the hamiltonian operator, the temperature average
of the fluctuation correlations depend on the quasi-particle occupation 
numbers $\nu_{k}$, $\langle \Pi^{'}_{{\bf k}} \Pi^{'}_{-{\bf k}} \rangle =$
$\; \langle \Phi^{'}_{-{\bf k}} \Phi^{'}_{\bf k} \rangle =$
$\; (1+2\nu_{k})/2$:
\begin{equation}
\frac{\partial F}{\partial \alpha_{k}}
= \frac{\partial}{\partial \alpha_{k}}
\left[ \left[ \frac{1}{2} (\alpha^{2}_{k} + \alpha^{-2}_{k}) (k^{2}/2m)
+ \mu \alpha^{2}_{k} \right] \; \frac{(1 + 2 \nu_{k})}{2} \; \right]
\; = 0 \; ,
\label{e:alpha}
\end{equation}
although the occupation numbers cancel out in
optimizing $\alpha_{k}$ and we find
\begin{equation}
\alpha^{2}_{k} = \frac{k^{2}/2m}{E_{k}} \; , 
\label{e:alpha2}
\end{equation}
where $E_{k}$ denotes the usual Bogoliubov dispersion relation,
$E_{k} = \sqrt{(k^{2}/2m+\mu)^{2} - \mu^{2}}$.
Unlike $\alpha_{k}$, expectation values of observable  quantities
do depend on the 
quasi-particle occupation numbers 
$\nu_{k} = \left( \exp[\beta E_{k}] 
-1 \right)^{-1}$.
In the last line of
(\ref{e:alpha2}) we display the square of the scaling parameter,
because $\alpha^{2}$ is the quantity that appears in the expression for the
dynamical structure factor, $\langle \hat{\rho}_{-{\bf q}} 
\hat{\rho}_{\bf q} \rangle
\approx n_{0}^{2} \delta_{{\bf q},0} +n_{0} \langle \Phi_{-{\bf q}} 
\Phi_{\bf q} \rangle$ $\; = n_{0}^{2} \delta_{{\bf q},0}
+ n_{0} \alpha_{q}^{2} \langle \Phi^{'}_{-{\bf q}} \Phi^{'}_{\bf q} \rangle$.  
Consequently, we find
for the dilute homogeneous BEC of low depletion,
the following expression for the dynamical structure factor density:
\begin{eqnarray}
\sigma^{({\rm hom})}_{\mu} ({\bf q},\omega ) &\approx&
\alpha^{2}_{q} n_{0} \; (2\pi)^{-1} \int dt'
\exp(i\omega t') \langle \Phi^{'}_{-{\bf q}} (t') \Phi^{'}_{\bf q} (0) \rangle
\nonumber \\
&=& \frac{q^{2}/2m}{E_{q}} \; n_{0} \left[
(1+\nu_{q}) \delta (\omega - E_{q}) + \nu_{q} \; \delta (\omega + E_{q})
\right] \; , 
\label{e:sigm}
\end{eqnarray}
which is the low-temperature generalization of the well-known zero-temperature
result \cite{Pines2}.  Note that the dependence on the occupation numbers
of (\ref{e:sigm}) is reminiscent of stimulated and spontaneous photon emission.

\section{Thomas-Fermi Dynamical Structure Factor}
	We calculate the structure factor $S({\bf q},\omega )$ in the 
quasi-homogeneous approximation (\ref{e:tfs}), using the above derived result
(\ref{e:sigm}) for the homogeneous BEC.  In this manner we find
\begin{eqnarray}
S_{TF}({\bf q},\omega) &=&
\int d^{3} R \; \; \sigma^{({\rm hom})}_{\mu ({\bf R})}
({\bf q},\omega)  
\nonumber \\
&=& \int d^{3} R \; n_{0}({\bf R}) \; \frac{q^{2}/2m}{\omega}
\left[ \left[ 1+\nu_{q}({\bf R}) \right] \delta 
\left(\omega - E_{q}({\bf R}) \right)
- \nu_{q}({\bf R}) \delta \left(\omega + E_{q}({\bf R})\right) \right] \; ,
\label{e:stf2}
\end{eqnarray}
where the position dependence of $\omega_{q}({\bf R})$ and $\nu_{q}
({\bf R})$, defined as
\begin{eqnarray}
&& E_{q}({\bf R}) = \sqrt{ \left( q^{2}/2m + \mu ({\bf R}) \right)^{2}
- \mu^{2}({\bf R})} \; \; ,
\nonumber \\
&& \nu_{q}({\bf R})
= \frac{1}{\exp \left[ \beta E_{q}({\bf R}) \right] -1} \; ,
\label{e:eo}
\end{eqnarray}
stems from the ${\bf R}$-dependence of the effective chemical potential
$\mu ({\bf R})$. We remark 
that in the quasi-homogeneous description, 
the phonon-like (or collective mode-like) delta-peak in the 
spectrum (\ref{e:stf2}), 
implies that the 
energy and momentum transfer, $\omega$ and $q$,
determine the spatial condensate region 
that is probed: the positions ${\bf R}$ for which $E_{q}({\bf R}) =
\omega$.

	We perform the spatial integration 
of Eq.(\ref{e:stf2}) 
for the simple example
of a spherically symmetric harmonic oscillator trap with potential $V(R)$,
\begin{equation}
V(R) = \frac{\omega_{T}}{2} (R/L)^{2} \; ,
\end{equation}
where $\omega_{T}$ is the trap
frequency and L the extent of its single-particle ground-state,
$L = 1/\sqrt{m \omega_{T}}$. 
The Thomas-Fermi expression for the condensate density in the low-depletion 
limit is easily obtained from the homogeneous result, 
$\lambda n_{0} \approx \mu$ (\ref{e:mu}), which gives 
the following quasi-homogeneous expression 
for the condensate density $n_{0}({\bf R})$:
\begin{equation}
n_{0}({\bf R}) = \frac{\mu ({\bf R})}{\lambda} =
\frac{\left[ \mu_{T} - V({\bf R}) \right]}{\lambda} \theta (R-R_{0}) \; . 
\end{equation}
The condensate radius $R_{0}$ ($ \; V(R_{0}) = \mu_{T}$, or 
$R_{0} = L \sqrt{2\mu_{T}/\omega_{T}} \; $),
is determined from the condition that the integral of the density is equal to
N, $R_{0} = L (15aN/L)^{1/5}$, where we have 
neglected the depletion \cite{usa}.
Spherical symmetry reduces
the integral (\ref{e:stf2}) to a one-dimensional 
integral over the radial distance R.
Finally, we substitute $R$ by the effective chemical
potential $\mu$,
\begin{equation}
R = R_{0} \; \sqrt{1-\mu /\mu_{T}}.
\end{equation}
In carrying out this substitution, we replace $n_{0}$ by $\mu /\lambda$ and
the delta functions by
\begin{equation}
\delta (\omega \pm E_{q}) \rightarrow \delta
\left(\mu - \mu_{q}(\omega ) \right)
\left| \partial E_{q} / \partial \mu \right|^{-1},
\end{equation}
where $\mu_{q}(\omega )$
is the effective chemical potential in the region where $E_{q}$ or
$-E_{q}$ equals $\omega$,
\begin{equation}
\mu_{q}(\omega ) = \frac{1}{2} \left[ \frac{\omega^{2}}
{q^{2}/2m} - \frac{q^{2}}{2m} \right],
\end{equation}
and $\partial E_{q}/
\partial \mu = q^{2}/2m/E_{q}$.
The resulting expression for the dynamical structure factor is simple:
\begin{eqnarray}
S_{TF}({\bf q},\omega) =
\left\{ \begin{array}{ll}
\frac{1}{2 \omega_{\rm T}} \left(\frac{R_{0}^{3}}{a L^{2}} \right) \frac{\mu_{q}
(\omega)}{\mu_{\rm T}} \sqrt{1-\frac{\mu_{q}(\omega)}{\mu_{\rm T}}} 
\times \left( 1 + \frac{1}{\exp(\beta \omega ) -1} \right)
& \mbox{if $\omega > 0$ and 
$0 < \mu_{q}(\omega) < \mu_{\rm T}$} \nonumber \\
\nonumber                   \\
\frac{1}{2 \omega_{\rm T}} \left(\frac{R_{0}^{3}}{a L^{2}} \right) \frac{\mu_{q}
(\omega)}{\mu_{\rm T}} \sqrt{1-\frac{\mu_{q}(\omega)}{\mu_{\rm T}}} 
\times \left( \frac{1}{\exp(\beta |\omega |) -1} \right)
& \mbox{if $\omega < 0$ and
$0 < \mu_{q}(\omega) < \mu_{\rm T}$} \nonumber \\
\nonumber 
0 &  \mbox{otherwise} \; ,
\end{array}
\right.
\label{e:fin}
\end{eqnarray}
which is the main result of this paper.  In Fig. 1, we show the dynamical
structure factor as a function of energy transfer,
for a fixed scattering angle or momentum 
transfer $q$ corresponding to  
$q^{2}/2m = 0.1 \mu_{T}$.
The temperature of the condensate is different for each curve,
$k_{B} T = 0$ (dot-dashed curve), $0.3 \mu_{T}$ (dashed curve)
and  $0.5 \mu_{T}$ (solid line). 
Notice that the ratio of the intensities distributed over
positive and negative transfer energies, is sensitive to the temperature.
Indeed, (\ref{e:fin}) 
satisfies the principle of detailed balance: $S({\bf q},\omega)
= \exp(\beta \omega ) S(-{\bf q},-\omega )$, which is a general result
\cite{Pines}.  This temperature dependence implies that
scattering experiments can directly measure the temperature of the atomic-trap
condensates simply by comparing the scattered intensities at $\omega$ and
$-\omega$ energy transfer, $T = \frac{\omega / k_{B}}{ \ln
\left[ S({\bf q},\omega )/S(-{\bf q},-\omega) \right] }$ . 

\section{Discussion of Limits of Validity} 
	Although the results of this paper are limited to dilute
condensates of low depletion, the quasi-homogeneous description has a much
broader range of validity.  One necessary condition 
for the validity of the dynamical quasi-homogeneous description,
is the validity of the static Thomas-Fermi model.  In the low
temperature region discussed in this paper, this amounts to the
requirement that the size of the condensate exceeds the extent of the
ground state, $R_{0} >> L$, or equivalently, 
$\mu_{T} >> \hbar \omega_{T}$
(\cite{Dalf},\cite{us},\cite{Kag}).  Furthermore, approximating the 
structure density $\sigma ({\bf R};{\bf q},\omega )$ by the value
of the homogeneous system is only sensible if the spatial variations
of the BEC are imperceptible in the region probed by  
$\sigma ({\bf R};{\bf q},\omega )$, i.e. if  q $>$ $l^{-1}_{v}$, where
$l_{v}$ is the scale on which the condensate varies spatially.  For example,
for a harmonic oscillator trap, we could choose $l_{v} \approx R_{0}/3$
since $\mu({\bf R})$ varies by approximately 10 \%  from the middle of
the trap to $R = R_{0}/3$.  The resulting restriction, $q > l_{v}^{-1}$,
requires a scattering angle larger than the scattering angle for
coherent scattering $q \leq R_{0}^{-1}$, so that the quasi-homogeneous
model is only useful in describing incoherent scattering.
Finally, we note that long-time fluctations, even on short distances,
are affected by the finite size effects.   Roughly speaking, if a localized
perturbation creates an excitation that
lives long enough to propagate to a region of different
density and reflect back to the position of origin, then one
should describe the fluctuations in terms of discrete harmonics or
eigenstates of the finite system.  The quasi-homogeneous description
is valid either if the lifetimes of the excitations are sufficiently small 
so that
the reflected excitation is damped out, or, if the scattering data
probe the fluctuations only over time periods less than $t_{v}$,
the time needed for the reflected excitation to return.  The restriction
to short-time fluctuations can be achieved by using an energy resolution
for the transfer energy that is less than or equal to 
$\Delta \omega \sim t_{v}^{-1}$. 
We estimate $t_{v}$ by assuming that 
the wave front of the excitation propagates at the local velocity
of sound, c = $\sqrt{\mu /m}$, and
we equate $t_{v}$ to the time necessary for the wave front 
to travel a distance $l_{v}$, $t_{v} \sim l_{v}/c$.  In the middle of
the trap, $c = \omega_{T} \times R_{0}/\sqrt{2}$, so that 
$t_{v} \sim \frac{l_{v}/R_{0}}{\omega_{T}}$ and thus
$\Delta \omega \sim t_{v}^{-1}
\sim \omega_{T} \times (R_{0}/l_{v})$.  We conclude by stating that the 
above considerations indicate that
the Thomas-Fermi
dynamical structure factor, as a function of the energy transfer,
should be interpreted as a `smooth' version of the
measured structure factor and that 
we should compare intensities integrated over frequency
intervals larger than or equal to $\Delta \omega$.
In reality, the estimate of $\Delta \omega$ 
depends on the region of the condensate that is probed, which in turn 
is determined by the value of the energy and momentum transfer.
The middle of the trap, which
is probed on the high--frequency side for $|\omega |$
($|\omega | \; \rm{near} \; \sqrt{(q^{2}/2m +
\mu_{\rm T})^{2} - \mu^{2}_{\rm T}} \;  $) 
requires the lowest value of
$\Delta \omega$ (estimated above), 
whereas on the low frequency-side, $|\omega | \sim q^{2}/2m$,
the edge of the condensate is probed, where the quasi-homogeneous description
cannot be trusted and $\Delta \omega \rightarrow \infty$.

\section{Acknowledgments}
	 P.T. was supported by Conselho Nacional de
Desenvolvimento Cientifico e Tecnologico (CNPq), Brazil.
The work of E.T. is supported by the NSF through a grant for the
Institute for Atomic and Molecular Physics at Harvard University
and Smithsonian Astrophysical Observatory.

\newpage

\underline{\Large Figure Caption}
\\
\\
\\
\underline{Figure 1}: Plots of the Thomas-Fermi dynamical structure factor
as a function of the energy transfer $\omega$, for fixed momentum transfer
$q$, $q^{2}/2m = 0.1 \mu_{T}$.  The three curves show the results at different
temperatures, $k_{B} T=0$ for the dot-dashed line, $k_{B} T = 0.3 \mu_{T}$ 
for the dashed curve, and $k_{B} T = 0.5 \mu_{T}$ for the plot shown in
solid line.

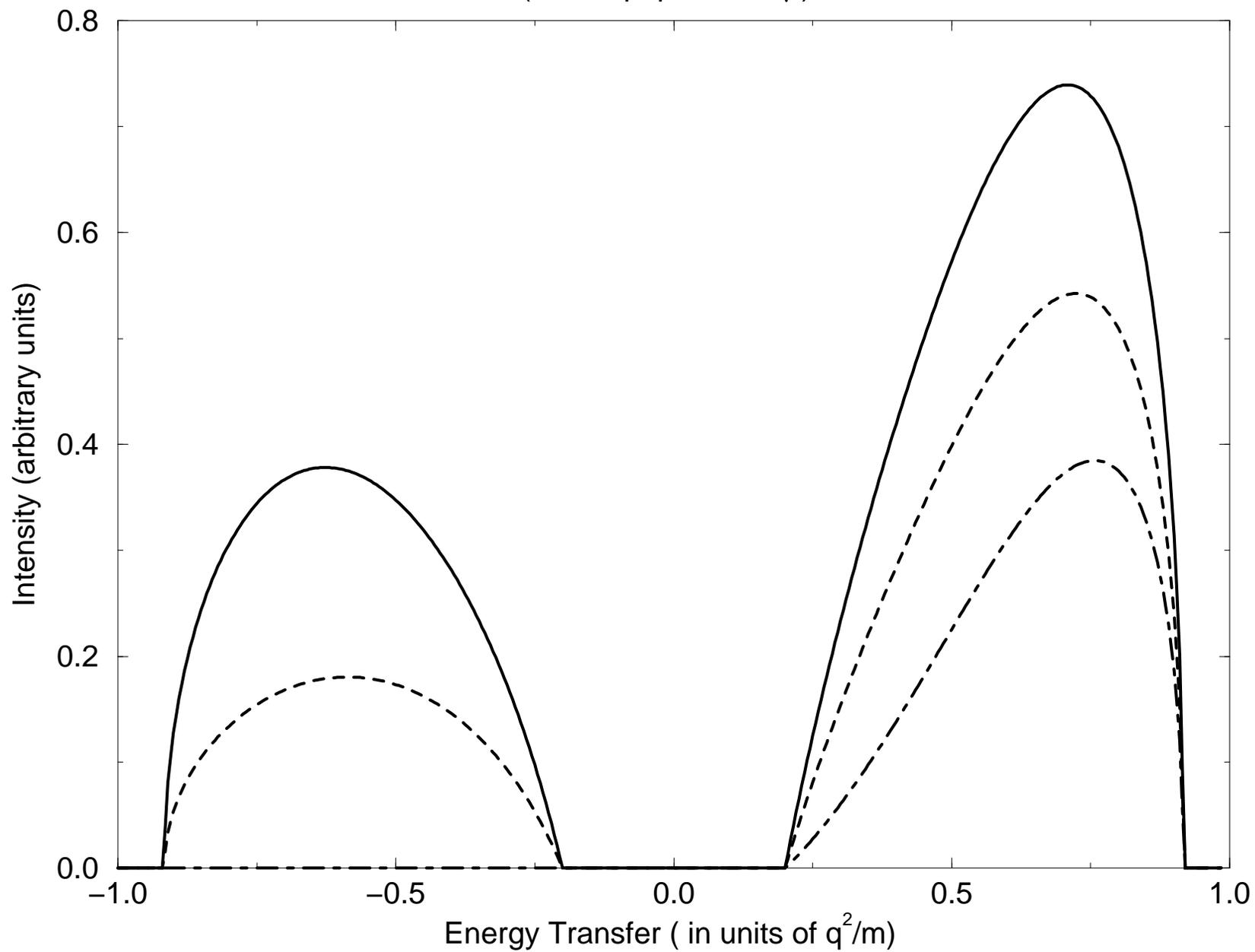
\end{document}